\documentclass[
 reprint,
 superscriptaddress,
 amsmath,amssymb,
 aps,
]{revtex4-1}

\usepackage{graphicx}% Include figure files
\usepackage{dcolumn}% Align table columns on decimal point
\usepackage{bm}% bold math
\usepackage{subfigure}
\begin{document}

\preprint{Arxiv}

\title{Films of bacteria at interfaces: three stages of behaviour}% Force line breaks with \\
%\thanks{A footnote to the article title}%
\author{Liana Vaccari}
\affiliation{Chemical and Biomolecular Engineering, University of Pennsylvania, Philadelphia, PA, 19104, USA}
\author{Daniel Allan}
\affiliation{Department of Physics and Astronomy, Johns Hopkins University, Baltimore, Maryland, 21218, USA}
\author{Nima Sharifi-Mood}
\affiliation{Chemical and Biomolecular Engineering, University of Pennsylvania, Philadelphia, PA, 19104, USA}
%\author{Lu Yao}
%\author{Nima Sharifi-Mood}
%\tnotetext[mytitlenote]{Fully documented templates are available in the elsarticle package on \href{http://www.ctan.org/tex-archive/macros/latex/contrib/elsarticle}{CTAN}.}
%\fntext[myfootnote]{These authors contributed equally to this work}
\author{Aayush Singh}
\affiliation{Chemical and Biomolecular Engineering, University of Pennsylvania, Philadelphia, PA, 19104, USA}
\author{Robert Leheny}
\email{leheny@pha.jhu.edu }
\affiliation{Department of Physics and Astronomy, Johns Hopkins University, Baltimore, Maryland, 21218, USA}
\author{Kathleen Stebe}
\email{kstebe@seas.upenn.edu}
\affiliation{Chemical and Biomolecular Engineering, University of Pennsylvania, Philadelphia, PA, 19104, USA}
%\cortext[mycorrespondingauthor]{Corresponding author}
%\ead{kstebe@seas.upenn.com}

%\date{\today}% It is always \today, today,
             %  but any date may be explicitly specified

\begin{abstract}
Bacterial attachment to a fluid interface can lead to the formation of a film with physicochemical properties that evolve with time. We study the time evolution of interface (micro)mechanics for interfaces between oil and bacterial suspensions by following the motion of colloidal probes trapped by capillarity to determine the interface microrheology. Initially, active bacteria at and near the interface drive superdiffusive motion of the colloidal probes. Over timescales of minutes, the bacteria form a viscoelastic film which we discuss as a quasi-two-dimensional, active, glassy
system. To study late stage mechanics of the film, we use pendant drop elastometry. The films, grown over tens of hours on oil drops, are expanded and compressed
by changing the drop volume. For small strains, by modeling the films as 2D Hookean solids, we estimate the film elastic moduli, finding values similar to those reported in the literature for the bacteria themselves. For large strains, the films are highly hysteretic. Finally, from wrinkles formed on highly compressed drops, we estimate film bending energies. The dramatic restructuring of the interface by such robust films has broad implications, e.g. in the study of active colloids, in understanding the community dynamics of bacteria, and in applied settings including bioremediation. 
\end{abstract}
\pacs{Valid PACS appear here}% PACS, the Physics and Astronomy
                             % Classification Scheme.
%\keywords{Suggested keywords}%Use showkeys class option if keyword
                              %display desired
\maketitle
\section{Introduction}
There is intense interest in the dynamics of microbes in suspensions, as they can be considered as active colloids. Bacteria in planktonic states convert chemical energy to propel themselves, and in dense suspensions, display rich collective behavior \cite{Goldstein,beer1,Kessler}. In this context, swimming microbes have inspired studies of reductionist systems in which particles convert chemical fuel into motion \cite{Sen, Howse}, which are analyzed in terms of non-equilibrium statistical mechanics \cite{Marchetti}. 
However, microbes are not merely colloids that move; they are living entities with complex surface chemistries and structures, which allow them to form complicated networks and to respond to external stimuli. Microbes also produce and secrete a variety of materials ranging from small molecule surfactants, polysaccharides, and growth factors that allow them to signal each other and to exist in communities \cite{Keller}. 
Via these secretions and associations, bacteria restructure their physicochemical environment.\\
The most common form of communal behavior is biofilm formation; an estimated $99$\% of the world's bacteria are sequestered in biofilms in various stages of growth \cite{Garrett,Dalton}. Biofilms comprise adherent cells in a complex three-dimensional matrix of extracellular polymeric substances\cite{Watnick} which resemble a disordered colloidal suspension in a polymer solution or gel\cite{Wilking}. These complex multicellular 3D macrostructures on solid surfaces grow in stages; microbes adhere to a surface, and grow into microcolonies that eventually form thick films \cite{OToole}. To form biofilms, microbes exploit synergistic interactions between cells coordinated by signaling via secreted enzymes and proteins \cite{Kolter}. In nature, the myriad roles played by biofilms continue to be identified. For example, biofilm formation can be beneficial for community survival, as biofilms protect the growing population of microbes against potential physicochemical and biological attacks \cite{Kimberly,Kokare}. Biofilms can protect bacteria from antimicrobial agents, with implications in antibiotic resistance  \cite{Dalton,Flemming}, and can confer evolutionary advantages via cooperation of co-colonized bacteria strains \cite{Griffin}. These protective mechanisms can be compromised, however, as motile bacteria of competing species can exploit the permeable, open structure of biofilms to invade and displace an established biofilm colony \cite{Houry}.\\
In more applied settings, biofilms play important roles. They can be remarkably strong elastic structures, and grow readily on bounding surfaces of pipes, reactors, ships, etc. As such, they decrease the efflux from factories, foul filters and bioreactors \cite{Goosen}, and are present in drag-increasing biocolonies on ships \cite{Schultz}. Biofilms can also be beneficial; for example, in bioremediation, bacteria and their biofilms remove toxins from their surroundings \cite{Flyvbjerg,Malik}.\\
The formation, structure, and mechanics of biofilms supported on solid surfaces are relatively well studied \cite{Wilking,Stoodley,Klapper,Rogers}. In contrast, films formed by microbes trapped at fluids interfaces have been less explored. We refer to such films involving bacterial cells as \textit{f}ilms of \textit{b}acteria at \textit{i}nterfaces (FBI). The nature of FBI and their relationship to the better-studied solid supported biofilms remain to be elucidated in terms of FBI composition, dynamical behavior and biological roles. There are a few notable recent studies in this field; for example, at air-aqueous interfaces,
the structure and properties of interfacial biofilms or pellicles of \textit{B. subtilis} \cite{Romero,Hobley} have been studied to investigate the role of amyloid fibers in providing structural integrity to the films. The time evolution and mechanics of FBI formed by {\textit{E. coli}} have been also studied, revealing rich interface rheology dependent on microbial secretions\cite {Aggarwal,Wu,Fischer,Fischer2,Fuller} with implications in infection \cite {Wu, Fuller}. FBI at oil-water interfaces have been studied \cite{Kang1,Kang2,Fischer,Abbasnezhad,kk11,Zoueki,Zoueki2}, motivated by their relevance to petroleum technologies, including oil recovery from oil sands \cite{Kang1,Kang2,Abbasnezhad,kk11,Zoueki,Zoueki2}. Such studies are also relevant to bioremediation, e.g. in the context of oil-consuming bacteria that appear near oil spill sites. FBI at oil-water interfaces form elastic skins, with film elasticities that depend on the surface structures of the microbes \cite{Wu2,Kang1,Kang2}. \\
Here we undertake a comprehensive study of the dynamic and spatial dependence of FBI formation in terms of film (micro)mechanics. We study FBI formation at initially bacteria-free hexadecane-aqueous interfaces for the model organism {\textit{Pseudomonas}} sp. ATCC 27259, strain $P62$. Our experiments reveal three stages of behavior: active films, viscoelastic films and elastic films. Initially, bacteria near the interface can be highly motile, forming an ÒactiveÓ layer. Over time, a film of adherent bacteria trapped in the interface forms; in the absence of nutrient addition, these bacteria are typically non-motile. At this stage, the FBI is viscoelastic. Particle-tracking experiments, in which film properties are revealed by trajectories of colloidal probes trapped at the interface, provide information about these stages.\\ 
To our knowledge, this is the first study that captures all of these stages, including the observation of active bacteria trapped at fluid interfaces and their transition to a viscoelastic film. The behavior of confined suspensions of motile bacteria near \textit{solids} has been intensely studied. Near solid boundaries, bacteria swim in recirculating patterns influenced by their confinement, their shape, and the nature of their bounding surfaces\cite{Wioland,Lushi08072014}. The emergence of these patterns is attributed to hydrodynamic interactions with shear flows\cite{Koser} and with image charges that enforce the no slip boundary condition\cite{Lauga2006400,Mehdi,Lauga}. We characterize the dynamics of our active layer by analyzing the motion of colloidal probes in the interface. Bacteria trapped at the interface by capillarity influence the motion of these probes. A population of non-adherent bacteria also influence the interface dynamics, leading to collective behavior which we compare to prior work in quasi-2D systems.\\
We infer that the viscoelastic transition is related to the secretion of polysaccharides and surfactants in the interface. This work complements existing studies in which secreted surfactants have been reported to play conflicting roles, as secreted surfactants are implicated in immobilization\cite{Zhang}, or in 
inducing superdiffusive motion at the air-water interface \cite{beer1}.\\
We address the FBI micromechanical properties from a fundamental perspective in the context of active, soft-glassy systems. FBI eventually form elastic, solid films. We probe the dilatational and bending moduli of these films using 
imposed area perturbations via pendant drop elastometry \cite{Knoche,Carvajal}, and provide evidence of nonlinear and hysteretic behavior for large area strains. These remarkably robust films significantly modify the mechanics of the interfaces on which they form. This restructuring of the physicochemical environment has broad implications in nature and in applied settings. 
\section{Experimental methods}
\subsection{Sample preparation}
\textit{Pseudomonas} sp. ATCC 27259, strain P62, chosen as a model organism, are cultured in ATCC Medium: 3 Nutrient Broth in $24$ ppt Instant Ocean to simulate a middle-level marsh salinity. The cultures are grown on a table-top shaker at 150 rpm at room temperature for $24$ h to mid-to-late exponential phase. For the FBI characterization experiments, a bacteria suspension free of surface active proteins from the nutrient broth is prepared with a typical washing protocol\cite{Kang2,Rosenberg1980} by centrifuging the culture for $5$ minutes at $5000\times g$, decanting the nutrient broth or supernatant and re-suspending the pellet in Instant Ocean three times.\\
\subsection{Particle tracking}
The particle-tracking measurements are performed in a $1.0$-cm I.D. cylindrical vessel with an inner surface whose bottom half is aluminium and top half is Teflon. When the cell is filled to the appropriate level, the aluminium-Teflon seam pins the oil-water interface, creating a flat interface with no meniscus. The base of the cylinder rests on an untreated glass coverslip, sealed with silica vacuum grease. To begin each experiment, the cylinder is filled nearly to the seam with $0.5$ ml of Instant Ocean. A spreading solution containing charge-stabilized polystyrene spheres (Invitrogen) with radius $R_s=0.5$ $\mu m$ in a mixture of equal parts by volume water and isopropanol is prepared in advance and sonicated to disperse any colloidal aggregates. To introduce the colloids to the interface, a $10$ $\mu L$ droplet of the spreading solution is gently placed in contact with the Instant Ocean surface. The solution wets the surface, and the colloids, which are slightly denser than water but hydrophobic, disperse across the interface. Droplets of hexadecane are then promptly placed on this system to form a film approximately $2~mm$ thick which covers the colloids and the aqueous suspension. The age of the sample $t_a$ is measured from the instant of oil-aqueous interface formation. After each experiment, the cylindrical vessel is cleaned thoroughly by scrubbing and sonicating in Alconox soap solution, acetone, and isopropanol, and then rinsed repeatedly in deionized water. Control microrheology experiments of dilute nutrient broth were performed absent bacteria to confirm the evolution of the interface was due to bacteria and film development.
We observe the colloids at the interface using an upright bright-field microscope with a $50\times$ objective. A camera (Zeiss AxioImager M1m) records a $200$ $\mu m$ by $150$ $\mu m$ field of view at $60$ frames per second, which set the shortest lag time over which probe motion can be characterized. Probe trajectories are extracted from the video using a custom Python implementation \cite{dan_allan_2014_12255} of the widely used Crocker-Grier multiple-particle-tracking algorithm \cite{Crocker}. Static and dynamic errors in the particle tracking are taken into account to avoid distortion in the measurements of the particle trajectories \cite{Savin}. Trajectories are extracted from video segments short enough in duration (typically 1-2 minutes) that no apparent change in particle mobility owing to the evolving interfacial properties can be discerned. Typically 30-200 probes are in view at a time, constituting up to $0.3\%$ surface coverage.
\subsection{Pendant drop elastometry} 
Pendant drop elastometry is used to study the evolution of the tension of the oil-water interface as the FBI forms. Changes in tension are also recorded under controlled deformations to the film achieved by expanding or contracting the droplet. These stress-strain observation are compared to appropriate models to extract material properties. A droplet of hexadecane is formed, in contact with the bacterial suspension, at the tip of an inverted needle by injection from a syringe. The drop is positioned in the path of a beam of light which projects the drop silhouette onto a CCD camera. The tension $\gamma_0$ is determined by comparing the edge of the droplet to a numerical solution of the Young-Laplace equation for an isotropic interface $\Delta p=\gamma_0(\kappa_s+\kappa_\phi)$ where $\Delta p$ is the pressure jump across drop interface, and $\kappa_s$ and $\kappa_\phi$ are the meridional and parallel curvatures, respectively. This form for the Young-Laplace equation can be recast in dimensionless form in terms of a Bond number for the droplet $Bo=\frac{\Delta \rho g a^2}{\gamma_0}$, where $R_0$ is the radius of curvature at the drop apex, $\Delta \rho$ is the density difference between the drop and the external phase, and $g$ is the gravitational acceleration constant. A comparison of numerical solutions to this equation and digitized experimental drop edges allows the tension $\gamma_0$ to be determined for a given drop image.\\
This analysis is appropriate immediately after drop formation. In later stages, the film behaves as a thin elastic sheet at the interface. If this sheet is unstrained, the isotropic form of the Young-Laplace equation and the standard pendant drop tensiometry remain appropriate. If, however, this sheet is strained, the principal stretches $\lambda_s$ and $\lambda_{\phi}$ are not isotropic, nor are the tensions, given by $\tau_s$ in the meridional direction and $\tau_{\phi}$ in the parallel direction. In this case, the relevant form for the Young-Laplace equation is $\Delta p=\tau _s\kappa_s+\tau _\phi\kappa_\phi$ with the associated tangential stress balance $-\frac{\cos\psi}{r}{\tau_\phi}+\frac{1}{r}\frac{d(r\tau_s)}{ds}=0$, where $s$ is the arclength measured from the apex of the droplet, $\psi$ is the turning angle and $r$ is the radial location of the drop edge in cylindrical polar coordinates. The drop shape is then determined by the relevant constitutive equation that relates local tensions ($\tau_s$ and $\tau_{\phi})$ to local stretches ($\lambda_s$ and $\lambda_{\phi}$). For an axisymmetric droplet in the limit of small deformations, the film can be described by a 2D Hookean model, for which the elastic response is dependent on the two material parameters, the two-dimensional elastic modulus $K_{2D}$ and the two-dimensional Poisson ratio $\nu_{2D}$. To determine these parameters, the drop is first imaged in its isotropic unstrained state to determine $\gamma_0$. Thereafter, the drop is deformed from that state and its silhouette is captured using a CCD camera. The edge of the strained drop is compared to solutions of the anisotropic Young-Laplace equation (see, c.f. Carvajal \textit{et al.} \cite{Carvajal} and Knoche \textit{et al.} \cite{Knoche} for details). To do so, we write an objective function for the difference between the experimental and numerical profiles and minimize it in a first order Newton-Raphson scheme with respect to the location of the axis of symmetry of the drop, the Bond number in terms of the unstrained tension, and the material parameters $K_{2D}$ and $\nu_{2D}$. From this analysis, we determine the elastic modulus $K_{2D}$. We find, however, that this method is insensitive to the two-dimensional Poisson ratio $\nu_{2D}$. Finally, following Knoche \textit{et al.} \cite{Knoche}, we estimate the bending modulus for the film from the wavelength of wrinkles that form under strong compression according to ${E_B}\sim\frac{{{\tau _s}\Lambda _c^4}}{{16{\pi ^2}L_w^2}}$ where $\Lambda _c$ is the wavelength of the wrinkles at the onset of wrinkling and $L_w$ is the length of the wrinkles which appear on the neck of the drop.
\section{Results and discussions} 
The interface between hexadecane and the aqueous bacteria suspension evolves with age, changing measurably over a timescale of minutes. We discuss three qualitatively distinct stages of the FBI development during which it can be characterized as (i) active, (ii) viscoelastic, and finally, (iii) as a solid elastic film. The first two stages of development are investigated primarily through the particle-tracking experiments, while the third stage of biofilm mechanics is characterized using the pendant drop experiments.
\subsection{Particle tracking} 
{\bf{Tracer Motion at an Active Interface}}\\ Immediately following the formation of a fresh interface between oil and the bacteria suspension, bacteria were sometimes observed to associate with the interface and remain motile once attached. Bacteria attach to the interface in both end-on and side-on orientations. Often, bacteria at the interface were accompanied by a population of especially mobile bacteria just beneath the interface. In these cases, the colloidal motion at the interface was strongly affected by hydrodynamic interactions with the swimming bacteria at and near the interface and by direct collisions with those at the interface. Typically, the concentration of swimming bacteria embodied a 10\% area fraction of the interface - large enough to lead to collective motion such as swirling, which influenced colloidal motion. A substantial literature has addressed the mobility of colloidal probes in the presence of motile bacteria and other microbial swimmers both in bulk (3D) and in quasi-two-dimensional contexts \cite{Libchaber,Kessler,Wilson,Soni,Chen,Mallouk,Jepson,Gollub,Rushkin,Kurtuldu}. One motivation for studying the colloidal dynamics in suspensions of swimming microbes is their utility as model systems for investigating the non-equilibrium statistical mechanics of active complex fluids. In addition, the colloid motion provides insight into biomixing, the enhanced transport of nutrients and other material by active suspensions. Since film formation relies on the transport of polysaccharides and other constituents to and within the interface, such biomixing could be an important feature of the film development. Hence, we have characterized the colloidal motion during this active phase in some detail.\\
Fig.~1(a) shows the colloids' ensemble-average mean-squared displacement $\left\langle {\Delta {r^2}(t)} \right\rangle = {\left\langle {{{({{\bf{ r}}_i}(t' + t) - {{\bf {r}}_i}(t'))}^2}} \right\rangle _{i,t'}}$, where the average is over particles $i$ and time $t'$ during this active stage of FBI formation, and ${\bf{r}}_i(t')$ is the time-dependent position of the $i$th particle. For reference, the mean-squared displacement of colloids at the oil interface of pure Instant Ocean containing no bacteria is also shown. In the absence of bacteria, $\left\langle {\Delta {r^2}(t)} \right\rangle$ varies linearly with lag time $t$ indicating simple diffusion, $\left\langle {\Delta {r^2}(t)} \right\rangle = 4{D_0}t$, with a diffusion coefficient, ${D_0} = 0.122~{\raise0.5ex\hbox{$\scriptstyle {\mu {m^2}}$}
\kern-0.1em/\kern-0.15em
\lower0.25ex\hbox{$\scriptstyle s$}}$, consistent with the viscosities of water and hexadecane. The mean-squared displacement of the colloids at the active interface similarly varies linearly with $t$ at large lag times, but with an enhanced effective diffusion coefficient. At smaller lag times, the mean-squared displacement grows more rapidly than linearly, indicating superdiffusive motion. Such superdiffusive motion is a common feature of colloidal motion in microbial suspensions and signals temporal correlations in the forcing of the colloids due to hydrodynamic interactions with the swimmers. A simple model for these correlations ascribes to them a single characteristic correlation time $\tau$, so that the particle velocities have an exponentially decaying memory, $\langle {\bf{v}}_i(t') \cdot {\bf{v}}_i(t'+t) \rangle_{i,t'} \sim
\exp(-t/\tau)$ \cite{Libchaber}. Such velocity correlations lead directly to a mean-squared displacement of the form \cite{Kurtuldu}, 
\begin{eqnarray}
\left\langle {\Delta {r^2}(t)} \right\rangle = 4D[t + \tau ({e^{ - t/\tau }} - 1)].\label{decay_exp}
\end{eqnarray}
In the limit of short lag times, $t \ll \tau $, this form predicts ballistic motion, $\left\langle {\Delta {r^2}(t)} \right\rangle = \frac{{2D}}{\tau }{t^2}$; at large lag times it reduces to diffusive motion, $\left\langle {\Delta {r^2}(t)} \right\rangle = 4Dt$, with effective diffusion coefficient $D$. The solid line in Fig.~1(a) is the result of a fit using this form, which describes the data accurately and gives $D = 1.8 \pm 0.1{\raise0.5ex\hbox{$\scriptstyle {\mu {m^2}}$}
\kern-0.1em/\kern-0.15em
\lower0.25ex\hbox{$\scriptstyle s$}}$ and $\tau = 0.053 \pm 0.007s$. The value of $D$ in relation to the diffusivity in the absence of bacteria, ${\raise0.5ex\hbox{$\scriptstyle {D}$}
\kern-0.1em/\kern-0.15em
\lower0.25ex\hbox{$\scriptstyle D_0$}}\sim15$, indicates that biomixing strongly influences the interface during this early stage of film formation. We note further that this factor likely underestimates the enhancement in tracer mobility due to the swimming bacteria since, as described below, even at these early interface ages the incipient film imparts an interfacial viscosity that reduces thermal diffusivity.\\
The success of Eq.~(1) in capturing the form of $\left\langle {\Delta {r^2}(t)} \right\rangle $ is consistent with several earlier studies of colloidal motion in active microbial suspensions and hence indicates that the velocities of tracers in such suspensions indeed have exponentially decaying memory. However, we can interrogate these correlations more directly by defining an instantaneous direction of motion for each particle,
\begin{eqnarray}
{{\bf{n}}_i}(\tau ) = \frac{{{{\bf{r}}_i}(\tau ) - {{\bf{r}}_i}(\tau - \delta \tau )}}{{|{{\bf{r}}_i}(\tau ) - {{\bf{r}}_i}(\tau - \delta \tau )|}},
\end{eqnarray}
where $\delta \tau={\raise0.5ex\hbox{$\scriptstyle 1$}
\kern-0.1em/\kern-0.15em
\lower0.25ex\hbox{$\scriptstyle {60}$}}~s$ is the time between successive video frames and $i$ labels the particles, and by examining its time-time autocorrelation function,
\begin{eqnarray}
{\Phi _n}(t) = {\left\langle {{{\bf{n}}_i}(\tau + t) \cdot {{\bf{n}}_i}(\tau )} \right\rangle _{i,\tau }}
\end{eqnarray}
The significance of ${\Phi _n}(t)$ is that it quantifies the persistence in the direction of the colloids' trajectories. As shown in Fig.~1(b), ${\Phi _n}(t)$ is effectively zero at $t>0.15~s$, setting the typical time required for the colloids' direction of motion to randomize completely. Notably, ${\Phi _n}(t)$ does not decay monotonically but has a peak near $t=0.03~s$. We attribute this peak to a tendency for colloids to follow ÒU-shapedÓ trajectories on short time scales due to hydrodynamic interactions with bacteria swimming past in close proximity \cite{Gollub}, and hence to make negative contributions to ${\Phi _n}(t)$. At larger times, ${\Phi _n}(t)$ decays exponentially, as shown by the line in Fig.~1(b), which is an exponential fit to the data. The correlation time obtained from the fit is $0.075 \pm 0.01s$, which is in reasonable agreement with the characteristic time $\tau$ for velocity correlations implied by fitting $\left\langle {\Delta {r^2}(t)} \right\rangle$ with Eq.~\ref{decay_exp}. Thus, through ${\Phi _n}(t)$ we observe directly the nature of the correlated tracer dynamics in an active suspension inferred by the analysis of $\left\langle {\Delta {r^2}(t)} \right\rangle$.\\
While the colloids' Brownian dynamics at large lag times with diffusivity $D$ suggest that the suspension of swimming bacteria acts like a thermal bath with large effective temperature, several previous studies have emphasized that the statistical properties of the colloidal displacements differ from those expected for a system in thermal equilibrium \cite{Chen, Gollub, Kurtuldu}. These differences are apparent in the probability distribution function (PDF) for displacements at fixed lag time ${P_t}(\Delta x)$, where $\Delta x$ is the displacement along one direction. Fig.~2(a) shows ${P_t}(\Delta x)$ at $t=0.0167~s$, $t=0.15~s$ and $t=1.3~s$, lag times spanning the superdiffusive and diffusive behavior in $\left\langle {\Delta {r^2}(t)} \right\rangle$. The PDFs of particles undergoing thermal diffusion in equilibrium would be Gaussian. In all cases, the PDFs in the active, incipient film show clear deviations from Gaussian forms, with tails at large $|\Delta x|$ that signal enhanced probability of large displacements. Qualitatively similar non-Gaussian PDFs have been observed previously among tracers in microbial suspensions and have been associated with advection-enhanced large displacements due to hydrodynamic encounters between the colloids and swimmers \cite{Gollub,Kurtuldu}. To compare the form and magnitude of the non-Gaussian contributions to ${P_t}(\Delta x)$ at different lag times more closely, we plot in Fig.~2(b) the normalized PDFs with displacement normalized by the root mean-squared displacement. Remarkably, the normalized PDFs collapse onto a single lineshape, indicating that the distribution function maintains a self-similar form with increasing lag time. Such self-similarity is generally unexpected and implies particular attributes about the colloidal dynamics, including (i) that the fraction of colloids in the non-Gaussian population remains constant over the range of lag times probed, and (ii) that the Gaussian and non-Gaussian displacements grow as the same function of lag time, first superdiffusively at short lag times and then diffusively at longer lag times. Self-similar distributions with non-Gaussian tails were also observed among tracer displacements within bulk (3D) suspensions of the eukaryotic microorganism \textit{Chlamydomonas}\cite{Gollub}. However, in that case, $\left\langle {\Delta {r^2}(t)} \right\rangle $ displayed diffusive behaviour over the entire range of lag times probed. The collapse in Fig.~2(b) is particularly notable because the lag times span both the superdiffusive and diffusive regimes. In both the previous case of tracers among swimming \textit{Chlamydomonas} and in our case of tracers in an incipient FBI, the non-Gaussian contributions to the probability distribution function follow a Laplace distribution, so that the total PDF can be described as the sum of two parts,
\begin{figure} 
\centering
{\label{SD1}\includegraphics[width=0.45\textwidth]{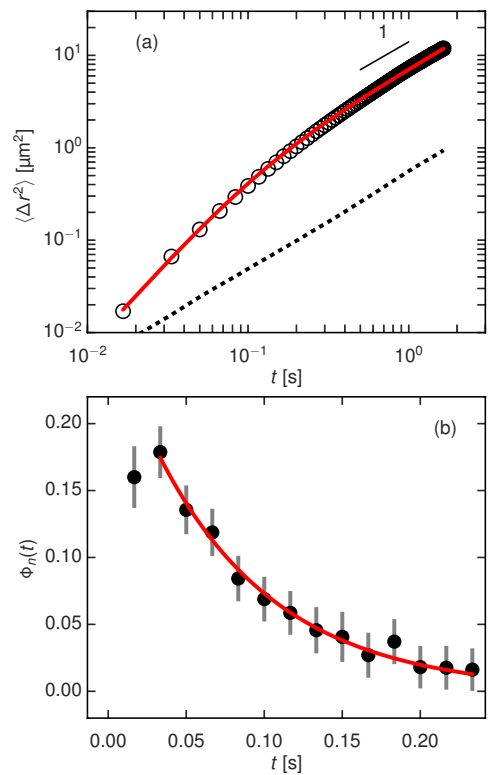}}\quad
\caption{\small{(a) Ensemble-average mean-squared displacement of colloidal tracers during the active stage of FBI formation at an oil-water interface. For reference, the dashed line displays the mean-squared displacement of colloids at the oil-water interface in the absence of bacteria. The solid line displays the result of a fit to the data with Eq.~\ref{decay_exp}. (b) Time autocorrelation function of the direction of instantaneous tracer displacements during the active stage of film formation. The line shows the result of an exponential fit to the data at large times ($t > 0.03~s$).
}}
 \label{SD}
\end{figure}
\begin{align}
{P_t}(\Delta x) = \frac{{1 - f}}{{{{(2\pi {\sigma ^2})}^{1/2}}}}\exp [ - \frac{1}{2}{(\frac{{\Delta x}}{\sigma })^2}] + \frac{f}{\xi }\exp ( - \left| {\Delta x} \right|/\xi ),\label{pdf_dis}
\end{align} 
where $f$ is the fraction of colloids in the non-Gaussian distribution, $\xi$ is the characteristic length of their displacements, and $\sigma$ is the width characterizing the displacements of the Gaussian distribution. The solid line in Fig.~2(b) is the result of a fit to the data using this form. The strong similarity between self-similar PDFs of tracer displacements in the incipient FBI and those in bulk suspensions of \textit{Chlamydomonas} is surprising since the tracer displacements in quasi-2D films of the \textit{Chlamydomonas} suspensions display qualitatively different non-Gaussian distributions, and the authors of those studies attributed the difference to the differing fluid velocity fields generated by the force dipoles of the \textit{Chlamydomonas} swimming in two and three dimensions \cite{Kurtuldu}. Our case of colloids entrained at an oil interface of a suspension of swimming bacteria that are forming a FBI has several features that distinguish it from these studies. First, the colloids, while in the incipient film, were in contact with the aqueous subphase and hence were coupled hydrodynamically to bacteria both in the film and in the bulk, a situation that is in some sense a hybrid of the three-dimensional and quasi-two-dimensional systems considered previously. Second, as discussed below, FBI formation substantially impacts the rheology of the interface even at the earliest ages, with possible qualitative consequences for the coupling between the swimming bacteria and the colloidal tracers in the film. Finally, \textit{Chlamydomonas} are ``pullers'' while \textit{Pseudomonas} are ``pushers'', and this distinction has consequences for the collective behavior and resulting hydrodynamics of suspensions. Given these distinctions, the strong correspondence between the PDFs in Fig.~2(b) and those from bulk \textit{Chlamydomonas} suspensions suggests that such self-similar distributions with Gaussian and Laplacian components might emerge more generically in non-equilibrium active systems than previously thought.\\
\begin{figure} 
\centering
{\label{SD1}\includegraphics[width=0.45\textwidth]{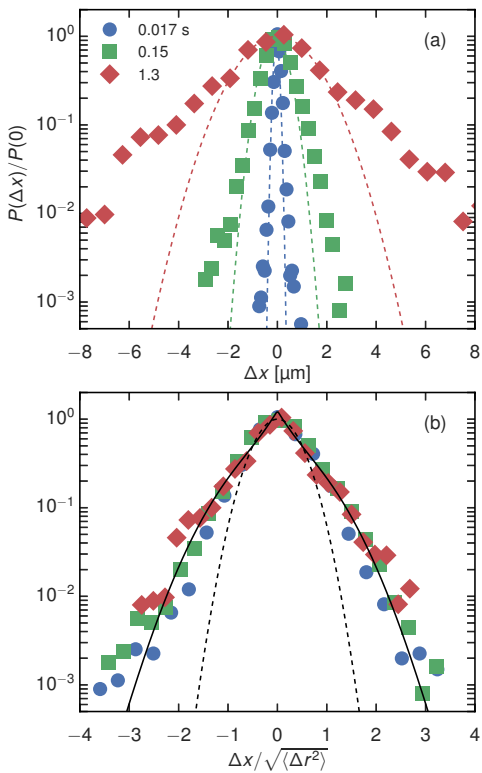}}\quad
\caption{\small{(a) Normalized probability distribution functions for colloidal displacements at three lag times during the active stage of FBI formation. The dotted lines display the results of fitting Gaussian distributions in the regions of the peaks of the distributions, highlighting the enhanced, non-Gaussian probability for large displacements. (b) Normalized probability distribution functions from (a) plotted against normalized displacement (scaled with the root mean-squared displacement at each lag time), illustrating the collapse of the distributions onto a universal curve. The dashed line displays the result of fitting a Gaussian distribution in the regions of the peak. The solid line displays the result of a fit using the form given by Eq.~\ref{pdf_dis}.}}
 \label{SD2}
\end{figure}
Another feature of our study of the active stage was the density of colloidal probes at the interface, which was large enough that correlated motion among the colloids could be observed, thereby providing information about the spatial correlations of the non-Brownian ``kicks'' the swimming bacteria impose. As an illustration, the inset of Fig.~3 depicts ${\bf{n}}$ for the colloids in the microscope field of view at one instant during the active stage. Alignment between the direction of motion of nearby colloids is clearly apparent. This coordinated motion is quantified in Fig.~3, which shows the normalized pair direction-direction correlation function,
\begin{eqnarray}
{C_n}(r) = {\left\langle {{{\bf n}_i}(\tau) \cdot {{\bf n}_j}}(\tau) \right\rangle _{i,j,\tau}}{\rm{ , }}
\end{eqnarray}
where the brackets represent an average over all pairs of particles $i,~j$ separated by distance $r$. The spatial correlations in tracer motion decay exponentially with separation, as depicted by the solid red line in Fig.~3, which is the result of an exponential fit to the data with correlation length of $4.7 \pm 1.7 \mu m$. Together, $G_n(t)$ and ${C_n}(r)$ give a quantitative picture of the short-time, spatiotemporal correlations in tracer motion that characterize the dynamical behaviour of the interface during the initial, active stage of FBI development.\\
{\bf{Viscoelastic Transition}}\\ Typically, the initial active stage of the film persisted for less than 5 minutes, after which no motile bacteria were observed either at the interface or in the near-interface bulk. We ascribe the limited duration of the bacteria motility to the lack of nutrient in the suspension. The end to the active stage was reflected in a qualitative change to the probe dynamics in which the probe mean-squared displacement changed from superdiffusive at short lag times to subdiffusive. Once the bacteria ceased to move visibly, we treated the interface as a passive system close to thermodynamic equilibrium and considered the probes to be undergoing thermally-driven Brownian trajectories from which the film rheology could be inferred.\\
\begin{figure}
\centering
\includegraphics[width=0.45\textwidth]{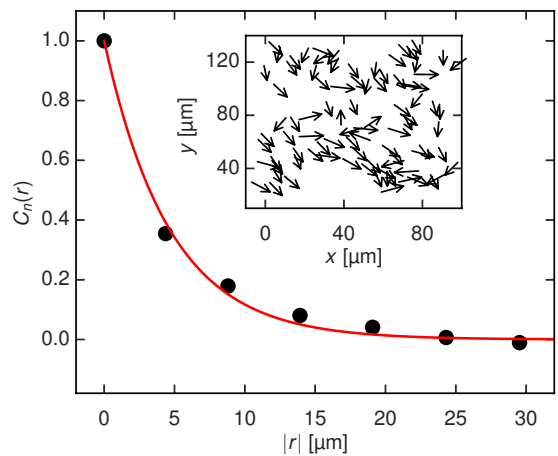}%
\caption{Normalized pair-direction correlation function as a function of the distance between colloids during the active stage of FBI development. Inset: Map of a section of the interface during the active stage of film formation showing the direction of motion of the colloidal tracers at an instant in time. Each colloid is represented by an arrowhead indicating its instantaneous direction of motion.}\label{figs}
\end{figure}
As mentioned above, the active stage was not always observed. Significantly, the ensuing mechanical changes of the interface, as inferred from probe mobility, appeared qualitatively independent of whether it was preceded by an active stage. Furthermore, as discussed below, the evolution in film rheology persisted for many minutes after the end of the active stage, suggesting that the presence of the active stage had limited impact on subsequent FBI evolution. From these observations we conclude that the film formation was primarily the consequence of polysaccharides and surface-active moieties produced by the non-motile (resting) bacteria. However, we cannot discount the possibility of some subtle effects to film formation due to biomixing by the swimming bacteria in those instances with an active stage.\\
Here, we present results for the viscoelastic evolution of the FBI from an experiment in which no protracted active stage was visible at the start of film formation. We focus on this data set because (i) the population of (non-motile) bacteria at the interface was limited, leaving a relatively unobstructed interface for the colloids and facilitating the analysis of their Brownian motion, and (ii) the colloidal motion was less likely to be affected by any residual activity than in a trial with a fully realized active stage. We emphasize again that the trends observed were qualitatively the same as those in trials in which the viscoelastic film development was preceded by an active stage.\\
Fig.~4(a) shows the ensemble-averaged mean-squared displacement $\left\langle {\Delta {r^2}(t)} \right\rangle $ of the colloidal probes at several ages $t_a$ of FBI formation following creation of the interface. Because each mean-squared displacement is determined from 1 minute of video, the maximum lag time $t$ is restricted to $t < 2s$ to assure adequate statistics. Again for reference, the mean-squared displacement of colloids diffusing at the oil interface of pure Instant Ocean¨ containing no bacteria is also shown. At our earliest measurement of film formation, the colloidal mean-squared displacement varies sublinearly with lag time, indicating a drag due to the development of a film with viscoelastic character.\\
Within this limited dynamic range of accessible lag times, $\left\langle {\Delta {r^2}(t)} \right\rangle$ is approximated as a power law, $\left\langle {\Delta {r^2}(t)} \right\rangle \sim{t^n}$ , with $n<1$. As shown in Fig.~4(b), the power-law exponent $n$ decreases steadily with increasing age, signifying increasingly subdiffusive motion. While in principle the probe mobility is affected by both the interfacial film and drag from the surrounding bulk oil and water, given the large difference between $\left\langle {\Delta {r^2}(t)} \right\rangle$ in the presence or absence of the forming FBI, we can safely infer that the bulk contributions to the drag are insignificant. In this case, under appropriate conditions one can obtain the frequency-dependent interfacial shear modulus, ${G^ * }(\omega ) = G'(\omega ) + iG''(\omega )$, from the Brownian motion of the probes through a two-dimensional version of a generalized Stokes-Einstein relation \cite{Helfer,T_Fischer},
\begin{eqnarray}
{G^ * }(\omega ) = \frac{{{k_B}T}}{{\pi i\omega {{\cal F}_u}\left[ {\left\langle {\Delta {r^2}(t)} \right\rangle } \right]}},
\end{eqnarray}
where ${{\cal F}_u}\left[ {\left\langle {\Delta {r^2}(t)} \right\rangle } \right]$ is the unilateral Fourier transform of the mean-squared displacement. Following Eq.~(6), power-law behaviour in the mean-squared displacements of the colloidal probes, $\left\langle {\Delta {r^2}(t)} \right\rangle \sim{t^n}$, with $n < 1$, implies the film's shear modulus has power-law frequency dependence $G'(\omega )\sim G''(\omega )\sim{\omega ^n}$ \cite{T_Fischer}. Such weak power-law frequency dependence of $G^{\ast}$ is a characteristic of the rheology of a broad range of disordered complex fluids including concentrated microgel solutions \cite{Ketz}, foams\cite{Khan}, paint \cite{Mackley}, intracellular matrix \cite{Fabry}, compressed emulsions \cite{Bibette}, clay suspensions \cite{Bonn}, and liquid-crystal nanocomposites \cite{Bandyopadhyay}, and is indicative of a broad spectrum of relaxation times. In most cases, the power-law exponent typically lies in the range $n \sim 0.1$ to $0.3$. The soft glassy rheology model explains this response as a general consequence of structural disorder and metastability, and provides a unifying theoretical framework for this behaviour. In this model, $n$ serves as an effective noise temperature, with systems approaching a glass transition as $n\to 0$. Thus, the steady decrease in $n$ with layer age reported in Fig.~4(b) points to increasingly glassy dynamics characterizing the structural response of the film.\\
\begin{figure} 
\centering
{\label{SD1}\includegraphics[width=0.45\textwidth]{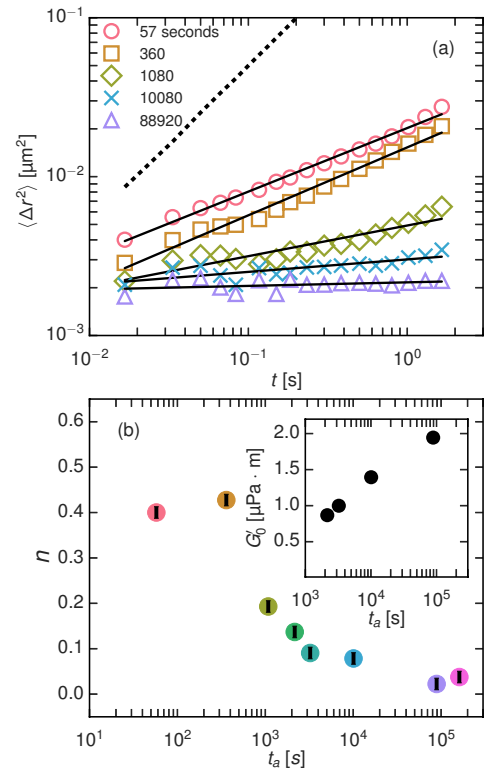}}\quad
\caption{\small{(a) Ensemble-average mean-squared displacement of colloids at several ages during FBI formation at an oil-water interface for a trial with no active stage. For reference, the dashed line depicts the mean-squared displacement of colloids at the bare oil-water interface in the absence of bacteria. The solid lines display the result of power-law fits. (b) Power-law exponent characterizing the ensemble average mean-squared displacements, $\left\langle {\Delta {r^2}(t)} \right\rangle \sim{t^n}$ , of colloids at the oil-bacteria solution interface as a function of the age since formation of the interface. Inset: The interfacial elastic shear modulus $G'_0$ at late ages, where the interface behaves like an elastic film. The elastic modulus grows logarithmically with age.}}
 \label{SD4}
\end{figure}
An important property of soft glassy systems is their non-equilibrium behaviour and spatial heterogeneity. As a measure of these features, Fig.~5(a)-(c) show the PDFs of colloidal displacements at three ages during the viscoelastic transition. In each case, the PDF is shown at three lag times normalized by the mean-squared displacement at that lag time. At the earlier two ages, ($t_a= 57$ and $2160~$s), during which the viscoelastic character of the film is evolving rapidly, the PDFs show a pronounced non-Gaussian component corresponding to enhanced probability of large displacements. These non-Gaussian contributions resemble those characterizing the colloidal dynamics in the active stage (Fig.~2); however, their origin in this case is different. Unlike at the active interface, the perturbations here are thermal, and each individual particle's displacements are Gaussian. The non-Gaussian distributions result from variation in the mobility of different particles, evincing a spatially heterogeneous interface of rheological microenvironments. Surprisingly, this heterogeneity diminishes at late ages, as illustrated by the closer-to-Gaussian distributions in Fig.~5(c), when the interface's evolution has slowed and the film is nearly elastic. An interesting future study would be to compare this spatial heterogeneity with that of films formed in the presence of an extended stage of activity by swimming bacteria to investigate the role of biomixing in suppressing such heterogeneity.\\
At late ages, when the layer behaves like an elastic film, and $\langle \Delta r^2 \rangle$ asymptotes to a constant value, we can extract an interfacial elastic shear modulus $G'_0$ from
\begin{equation}
G'_0 = \frac{k_B T}{\langle \Delta r^2(t \rightarrow \infty) \rangle}
\end{equation}
As shown in Fig.~4(b) inset, $G'_0$ grows logarithmically with layer age. Such logarithmic growth of the elastic modulus is a characteristic of the aging behavior of disordered soft solids such as colloidal gels and pastes~\cite{Derec2003,Coussot2006}, indicating again the similarity between the rheology of the biofilm and that of soft glassy materials.
\begin{figure}
\centering
\includegraphics[width=0.45\textwidth]{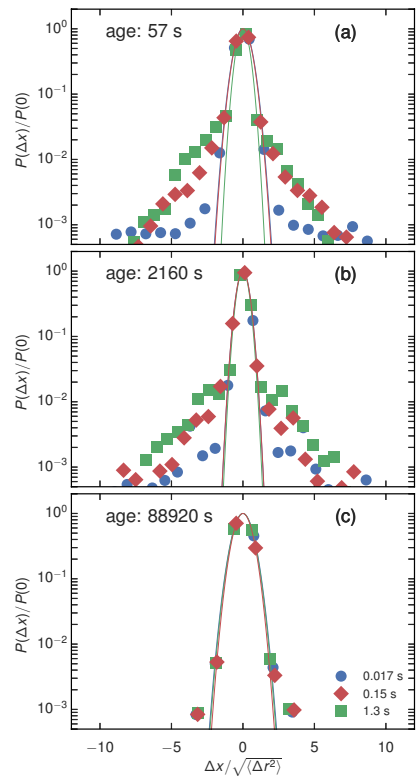}%
\caption{Normalized probability distribution functions for colloidal displacements at three lag times and three film ages (a) $57~s$, (b) $2160~s$, and (c) $88920~s$ during the viscoelastic stage of FBI formation. The solid lines display the results of fitting Gaussian distributions in the regions of the peaks of the distributions, highlighting the enhanced, non-Gaussian probability for large displacements.
}\label{figs1}
\end{figure}
\begin{figure}
\centering
\includegraphics[width=0.5\textwidth]{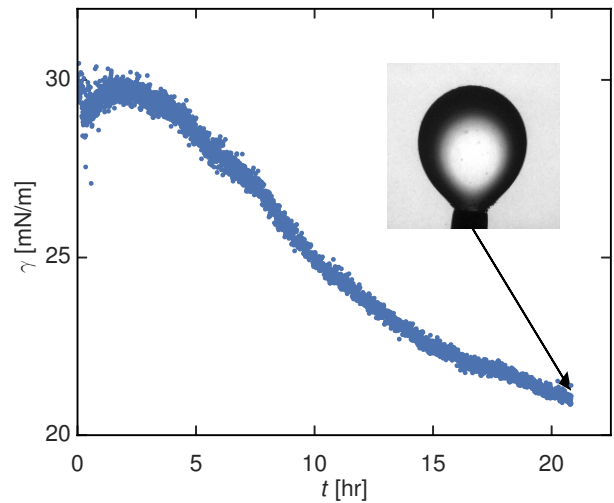}%
\caption{Dynamic surface tension of a hexadecane drop aged in a bacteria suspension. Inset image shows the drop shape at the end of the aging cycle. The needle is $0.9081~mm$ wide and serves as a scale bar.}\label{figs2}
\end{figure}
\subsection{Pendant Drop Elastometry}
A pendant drop of hexadecane is placed in contact with a bacterial suspension. As bacteria adhere to the drop interface, the FBI forms and the interfacial tension decreases (Fig.~6). In early stages, the colloidal probe data above indicate that the interface is fluid, and the decrease in interfacial tension is attributable to surface active polysaccharides secreted by adsorbed bacteria. Eventually, a film forms at the interface with significant elasticity. To probe this elasticity, the biofilm is deformed from its unstrained state by injecting or withdrawing liquid from the droplet, thereby changing the drop volume and area. Here we analyse a data set which captures key features of the FBI in late stages of development. A drop of hexadecane aged in a bacteria suspension for $22$ hours has an apparently smooth surface with a shape corresponding to that for a pendant drop subject to an isotropic tension of $\gamma_0=21.0~{\raise0.5ex\hbox{$\scriptstyle {mN}$}
\kern-0.1em/\kern-0.15em
\lower0.25ex\hbox{$\scriptstyle m$}}$. When the drop volume is decreased, the drop shape changes dramatically, as shown in Fig.~7. For small enough compressions (${\raise0.5ex\hbox{$\scriptstyle { - \Delta A}$}\kern-0.1em/\kern-0.15em\lower0.25ex\hbox{$\scriptstyle {{A_0}}$}}< 0.3$), the drop shapes are used to infer the two-dimensional elastic modulus $K_{2D}$ and the two-dimensional Poisson ratio $\nu_{2D}$ of the FBI by assuming a 2D Hookean constitutive model. For larger compressions, the drops wrinkle at their necks; this wrinkling phenomenon is discussed below. For still larger compressions, the drop resembles an empty, wet plastic sack with pronounced wrinkles. 
To analyse the small deformation regime, we integrate the anisotropic Young-Laplace equation for 2D Hookean films with unstrained tension corresponding to the apparent tension of the drop prior to compression, $\gamma_0=20.44~{\raise0.5ex\hbox{$\scriptstyle {mN}$}\kern-0.1em/\kern-0.15em \lower0.25ex\hbox{$\scriptstyle m$}}$. Consider the tension, strain and shape profiles, shown in Fig.~8; in this figure, lengths are scaled with the radius of curvature of the unstrained drop $R_0 =1.30~mm$ and tensions $\tau_s$ and $\tau_\phi$ are scaled with $\gamma_0$. The shapes of the unstrained (dashed curve) and compressed drops (solid curve) are shown in Fig.~8(a). The apex of the compressed drop is located at the origin; this contour ends where the drop would intersect the needle at the right. The meridional tension $\tau_s$ and parallel tension $\tau_\phi$ profiles are shown in Fig.~8(b), for ${\raise0.5ex\hbox{$\scriptstyle { - \Delta A}$}\kern-0.1em/\kern-0.15em\lower0.25ex\hbox{$\scriptstyle {{A_0}}$}} = 0.13$ for material parameters that capture the experimental drop profiles well, i.e. ${{{K}}_{{{2D}}}} = 21.3~{\raise0.5ex\hbox{$\scriptstyle {mN}$}\kern-0.1em/\kern-0.15em\lower0.25ex\hbox{$\scriptstyle m$}},~\nu_{2D} {\rm{ = 0.78}}$. For much of the contour, the tensions are equal but non-uniform; they are highest at the apex where strains are weakest, and decrease steeply where the drop neck meets the needle. The high strain in this region is consistent with the area dilatation profile (Fig.~8(c)). This anisotropic strain environment is key to the formation of wrinkles in the layer at large compressions \cite{Knoche}. In order for the film elasticity to give pronounced changes in drop shape like those evident in our experiments, ${K_{2D}}$ must be similar in magnitude to $\gamma_0$. {{Values for this elastic modulus are obtained by minimizing an objection function, which is the cumulative distance between edge points on the experimental drop and their respective nearest neighbor on the numerical drop, with respect to Bond number, $Bo$, location of the axis of symmetry, radius of curvature at the apex, $R_0$, elastic modulus $K_{2D}$, and Poisson ratio, $\nu_{2D}$. Examples of an intermediate fit and a fully optimized numerical drop contour are shown in Fig.~9. The resulting data, (Fig.~10), show $K_{2D}$ of a characteristic magnitude, in the range of $10-27~{\raise0.5ex\hbox{$\scriptstyle {mN}$}\kern-0.1em/\kern-0.15em\lower0.25ex\hbox{$\scriptstyle m$}}$, with a downward trend suggesting that this is not a Hookean film in this strain regime. This modulus is far larger than that found for hexadecane drops emulsified and aged over minutes in the presence of bacteria characterized by conventional pendant tensiometry and micropipette aspiration \cite{Kang1,Kang2}, for which a modulus of $5~{\raise0.5ex\hbox{$\scriptstyle {mN}$}\kern-0.1em/\kern-0.15em\lower0.25ex\hbox{$\scriptstyle m$}}$ is reported. Pendant drop elastometry assuming isotropic films was performed for bacteria grown on mineral and medium chain triglyceride oils, with elasticities ranging from $10-30~{\raise0.5ex\hbox{$\scriptstyle {mN}$}\kern-0.1em/\kern-0.15em\lower0.25ex\hbox{$\scriptstyle m$}}$ and $2-10~{\raise0.5ex\hbox{$\scriptstyle {mN}$}\kern-0.1em/\kern-0.15em\lower0.25ex\hbox{$\scriptstyle m$}}$ respectively \cite{Fischer2}. Interestingly, the modulus we find is comparable to the spring constants for individual planktonic gram negative bacteria as characterized by AFM, which range from $10-20~{\raise0.5ex\hbox{$\scriptstyle {mN}$}\kern-0.1em/\kern-0.15em\lower0.25ex\hbox{$\scriptstyle m$}}$; bacteria may form biofilms with elastic moduli comparable to their own elasticities 
\cite{Volle}.\\
Note that values for the 2D Poisson ratio extracted by the scheme are highly variable (inset in Fig.~10). This variation is not physically significant; we find that pendant drop shapes are very weakly dependent on this quantity. To illustrate this concept, a family of simulated profiles of Hookean pendant drops is shown (Fig.~11), generated from a single unstrained state and elastic modulus but differing 2D Poisson ratios. The profiles superpose for most of the drop contour, differing only in the high strain region of the drop neck. This is also the site of highest strain anisotropy where the role of the 2D Poisson ratio would be most significant. Well-resolved edge detection in the region of the neck would be extremely important to distinguish between these values; in many cases, in experiment, a droplet might attach to a needle before these curves peel away from each other, limiting the degree of precision with which one can evaluate the Poisson ratio. Furthermore, in our study, bacteria aggregates adhered to the FBI obscure the drop shape near the neck, precluding the relevant analysis, which rely on having a well resolved drop profile.\\ 
For large compressions, the neck region of the drop eventually roughens. Upon continued compression, the rough regions deepen into wrinkles which lengthen vertically near the neck. Such wrinkled elastic films formed by FBI have been reported previously \cite{Kang2,Hobley}.\\
The appearance of these wrinkles can be used to show that the 2D Poisson ratio of the film is positive. Furthermore, we can use these wrinkles to show the bending energy $E_B$ of the film. Under strong compression, tension reduces more in the parallel direction than in the meridional direction, which must support gravity's pull in the vertical direction. These anisotropic tensions and associated strains cause wrinkles to form aligned with the meridional direction. The onset of this deformation occurs when the energy for compression becomes comparable to the energy required to wrinkle, characterized by the bending energy $E_B$. At onset, wrinkles of a single wavelength $\Lambda _c$ and associated amplitude should appear for $\Lambda _c \ll R_0$ which is related to the bending energy according to ${E_B}\sim\frac{{{\tau _s}\Lambda _c^4}}{{16{\pi ^2}L_w^2}}$ where $L_w$ is the length of wrinkled region and $\tau_s$ is the meridional tension at onset \cite{Knoche,Erni}. In our experiments, when wrinkles become apparent, however, deep wrinkles with wide spacing coexist with wrinkles of smaller amplitude and shorter wavelengths. The deep, widely spaced modes appear more prominent as compression is increased (Fig. 7(b)). Strictly speaking, the emergence of this band of wavelengths indicates compression beyond the onset condition. However, to infer an order of magnitude for the bending energy, we associate the longest wavelength with $\Lambda_c$ (given as a range because of the decrease in drop radius along the arc length), and estimate $\tau_s$ from the last drop silhouette before wrinkling, to find an estimated range for $E_B\sim 2 - 50 \times {10^{ - 16}}~N \cdot m$ \cite{Knoche}. This range of values is consistent with those reported previously for cross linked films of polymers (OTS) at fluid interfaces \cite{Knoche,PhysRevLett.90.074302,Seifert}.\\
\begin{figure*}
\centering
{\label{SD1}\includegraphics[width=0.9\textwidth]{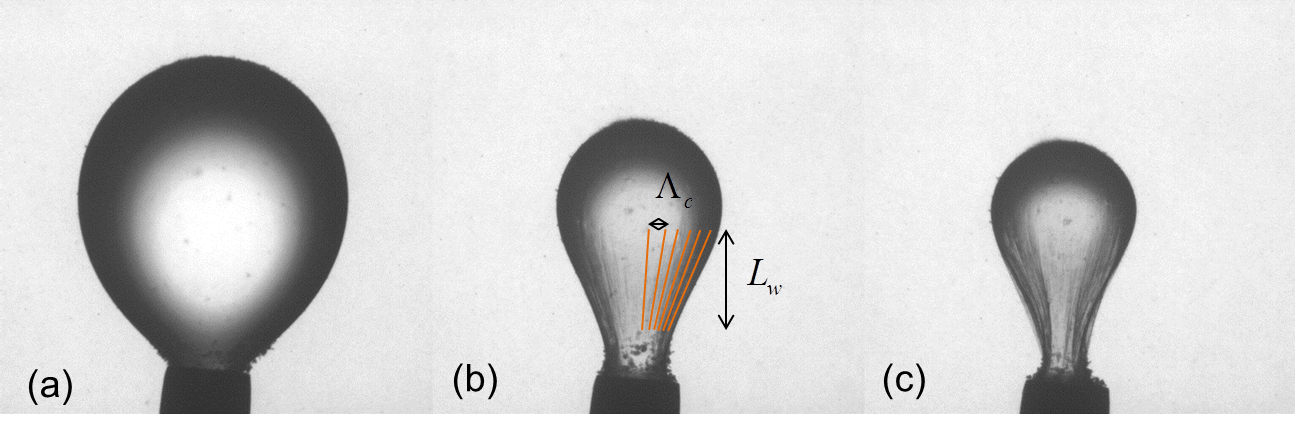}}\quad
\caption{\small{Compression of FBI aged for $22$ hours on a drop of hexadecane by withdrawal of oil. The compression increases from (a) to (c). The needle is $0.9081~mm$ wide and serves as a scale bar. As the drop is compressed, the film wrinkles. Approximate wrinkle analysis of (b), an intermediate stage of compression: $\Lambda_c\sim 5-12\times10^{-5}~m$, $L_w\sim1.1\times10^{-3}~m$, $\tau_s\sim5~~{\raise0.5ex\hbox{$\scriptstyle {mN}$}
\kern-0.1em/\kern-0.15em
\lower0.25ex\hbox{$\scriptstyle m$}}$.
}}
 \label{SD7}
\end{figure*}
Finally, we investigate the hysteresis of these films under compression and re-expansion. The FBI stiffen significantly even after a single compression/re-expansion cycle. Fig.~12 shows this phenomenon. Panel $A$ corresponds to the film-covered drop prior in its unstrained state. In this experiment, a pendant drop was aged in the bacterial suspension for 67 hours, compressed to wrinkling, and subsequently re-expanded. Panel $B$ is the drop after compression. Panel $C$ corresponds to the drop after re-expansion to its initial volume and area, and $D$ to areas well in excess of the initial drop (details are given in the figure caption). Had the film not been hysteretic, the drop would have attained the same shape at a given volume or area. In both panels $C$ and $D$, the drop is apparently more spherical, illustrating its higher film tension in each, clear evidence of hysteresis. The microstructural origins of this apparent increase in tension have yet to be determined; possible mechanisms include irreversible folding of the film where wrinkles formed, effectively reducing the film area and creating stiffer regions, or irreversible collapse of the porous microstructure of the film.\\
These encapsulating films provide a strong barrier to coalescence \cite{Kang1}, in which a drop of oil aged in a bacterial suspension for $22$ hours is compressed against a FBI formed at a planar oil-water interface. Significant compression is supported by the films when pressed against each other, with no evidence of rupture.
\begin{figure}
\centering
\includegraphics[width=0.45\textwidth]{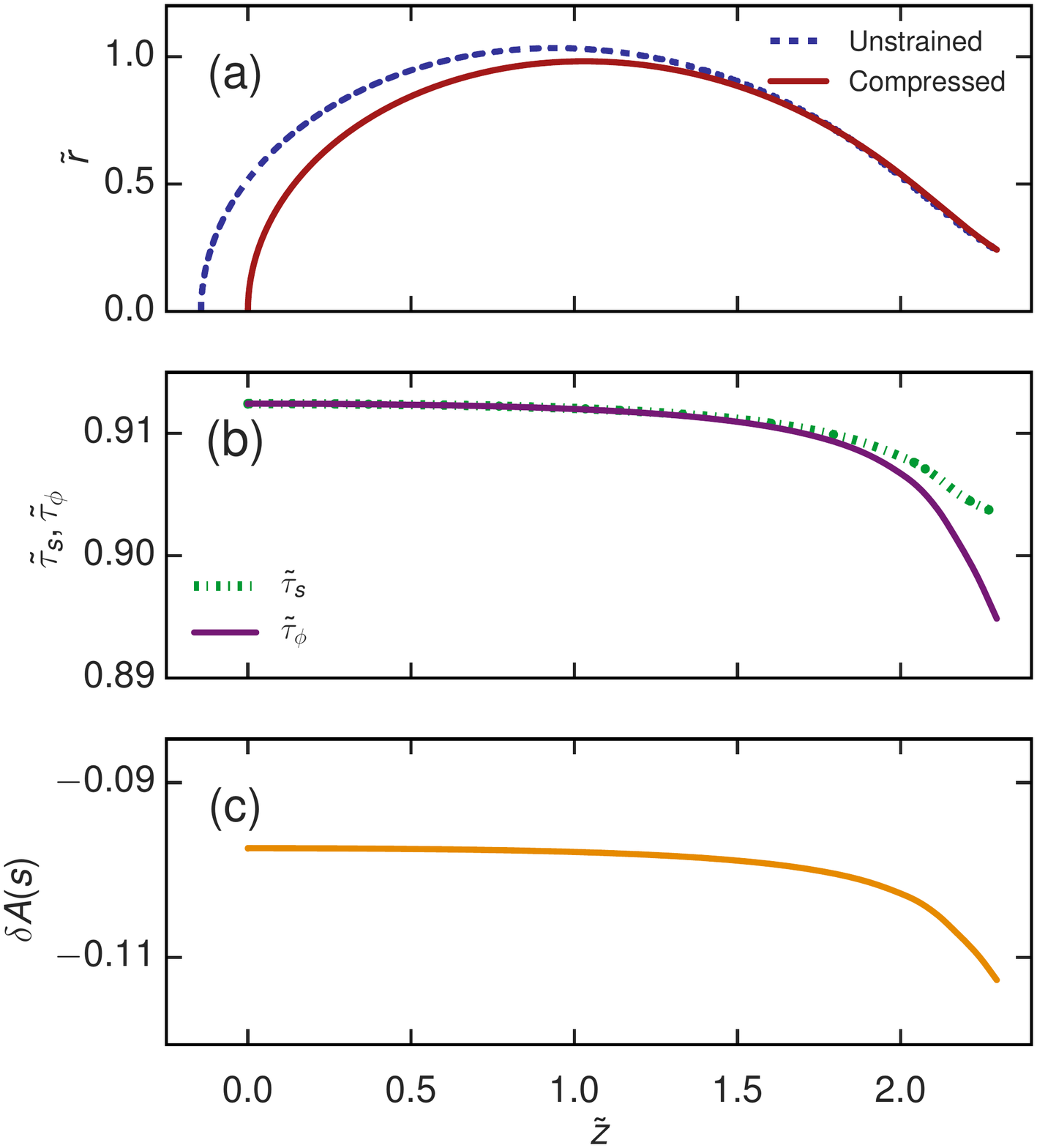}%
\caption{Non-dimensional (a) theoretical shape, where (0,0) is the apex of the drop, and the needle would be at the termination on the right, (b) principal tensions, parallel, $\tau_\phi$, and meridional, $\tau_s$, and (c) local area dilatational ($\delta A(s)$= $\lambda_{\phi}\lambda_s-1$) profiles of a drop under compression: $Bo = 0.18$, $\gamma_0 = 20.44 ~{\raise0.5ex\hbox{$\scriptstyle {mN}$}
\kern-0.1em/\kern-0.15em
\lower0.25ex\hbox{$\scriptstyle m$}}$, $K_{2D} = 17.0 ~{\raise0.5ex\hbox{$\scriptstyle {mN}$}
\kern-0.1em/\kern-0.15em
\lower0.25ex\hbox{$\scriptstyle m$}}$, $\nu = 0.78$, $R_a = 1.24~mm$, $V = 9.8~\mu l$.}\label{figs8}
\end{figure}
\begin{figure}
\centering
{\label{SD1}\includegraphics[width=0.45\textwidth]{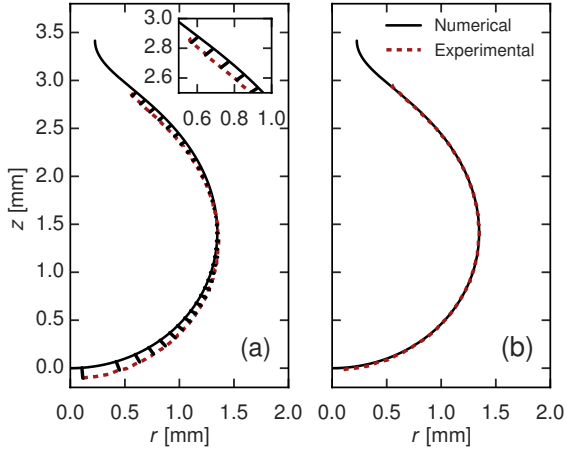}}\quad
\caption{{{Fitting of experimental drop shape to numerical anisotropic Young-Laplace equation solution. (a) Experimental contour and the numerically computed drop shape for an initial guess for the elastic modulus, $K_{2D}$, and Poisson ratio, $\nu_{2D}$. (b) Best fit numerical solution to experimental data.}}}
 \label{SD7}
\end{figure}
\begin{figure}
\centering
\includegraphics[width=0.45\textwidth]{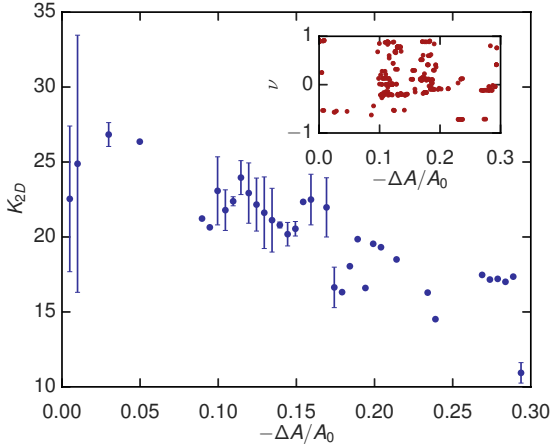}%
\caption{Anisotropic analysis of a compression cycle: elastic modulus over $60$ bins, with an average $K_{2D} = 20.81 ~{\raise0.5ex\hbox{$\scriptstyle {mN}$}
\kern-0.1em/\kern-0.15em
\lower0.25ex\hbox{$\scriptstyle m$}}$. Inset: 2D Poisson ratio varies across the whole range of possible values with no discernible trend.}\label{figs9}
\end{figure}
\begin{figure}
\centering
\includegraphics[width=0.49\textwidth]{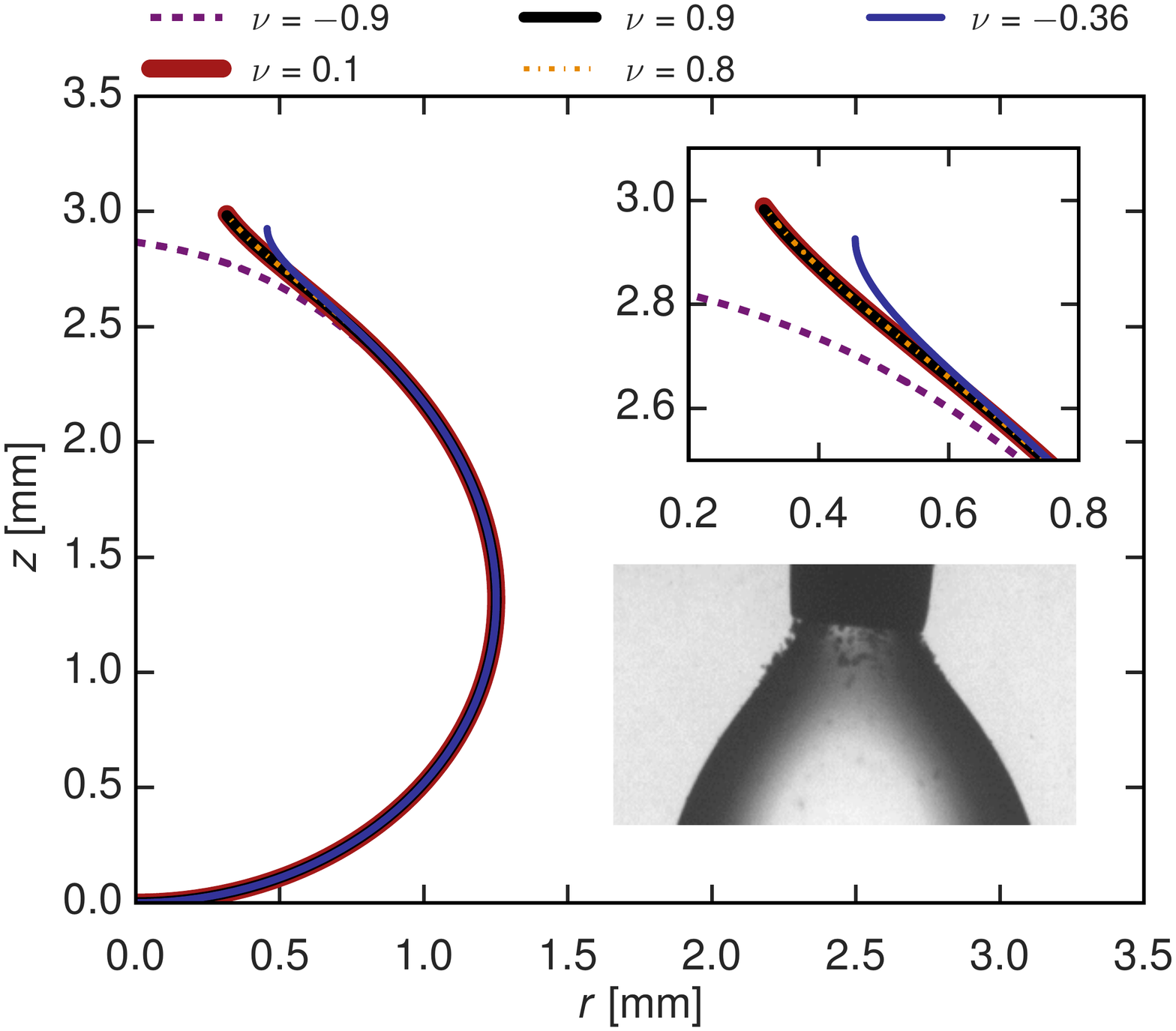}%
\caption{Illustration of weak dependence of pendant drop shapes on the Poisson ratio in the region of the drop neck on a drop with $Bo = 0.18$, $\gamma_0 = 20.44 ~{\raise0.5ex\hbox{$\scriptstyle {mN}$}
\kern-0.1em/\kern-0.15em
\lower0.25ex\hbox{$\scriptstyle m$}}$, $K_{2D} = 17.19 ~{\raise0.5ex\hbox{$\scriptstyle {mN}$}
\kern-0.1em/\kern-0.15em
\lower0.25ex\hbox{$\scriptstyle m$}}$, $R_a = 1.22~mm$. The needle is $0.9081~mm$ wide and serves as a scale bar. Inset plot shows an expanded plot of the family of $\nu$ curves at the neck. Inset image shows aggregation by bacteria at the neck of the experimental drop.}\label{figs10}
\end{figure}
\section{Conclusions}
We study the formation of films at interfaces of bacterial suspensions (FBI), capturing associated changes in (micro)mechanics. Films may exhibit an early, active stage in which hydrodynamic interactions with bacteria near and at the interface create superdiffusive motion of colloidal probes in which the probe velocity has an exponentially decaying memory. The probability distributions of probe displacements are non-Gaussian but are self-similar at different lag
times, even at times far exceeding the characteristic time of the velocity memory. The active stage is followed by a viscoelastic stage during which the bacterial suspension forms a heterogeneous film with micromechanics that evolve over time. The film rheological properties behave similarly to a broad range of disordered complex materials in that the storage and loss moduli scale with $\omega^n$ where $n<1$. Notably, the free interface displays remarkable changes in viscoelasticity after only a few minutes. This observation suggests that any study of the dynamics of bacteria suspensions with free interfaces must be treated with care, as collective dynamics depend sensitively on the mobility of bounding surfaces. Over time, the FBI become elastic via the formation of heterogeneous patches which eventually cover the interface. In this state, the film exhibits signature behaviour of thin elastic shells with associated bending energies and anisotropic tensions. We use pendant drop elastometry to obtain the properties of these films after more than 20 hours of interface aging. By weakly compressing and expanding a FBI-covered oil drop, we determine the elastic moduli, which are comparable to "spring constants" determined for Gram negative bacteria by AFM. By strongly compressing the FBI-covered oil drop, the films wrinkle, implying a positive Poisson ratio associated with the interfacial film. From the wave lengths of the wrinkles, we also extracted a bending energy consistent with literature values for thin polymer films. Additionally, these FBI exhibit significant hysteresis upon expansion and compression. \\
Finally, we have studied the behavior of a particular bacterium, {\textit{Pseudomonas}} sp. ATCC 27259, strain $P62$. Different species or strains have different propensities to form FBI; these propensities depend on the presence or absence of nutrient sources. Furthermore, for a given strain, film (micro)mechanics likely depend on the surface structures and motility of the bacteria. We will expand this work to address such issues, and to investigate community dynamics of competing and synergistic bacteria.
\section{Acknowledgements} 
Acknowledgment is made to Dr. Tagbo Niepa and Professors Daeyoen Lee, Mark Goulian and Jian Sheng for helpful discussions. Acknowledgement is also made to the Donors of the American Chemical Society Petroleum Research Fund and to GoMRI (Gulf of Mexico Research Initiative) under the grant No. SA12-03/GoMRI-003 for partial support of this research.\\
\begin{figure*}
\centering
{\label{SD1}\includegraphics[width=0.9\textwidth]{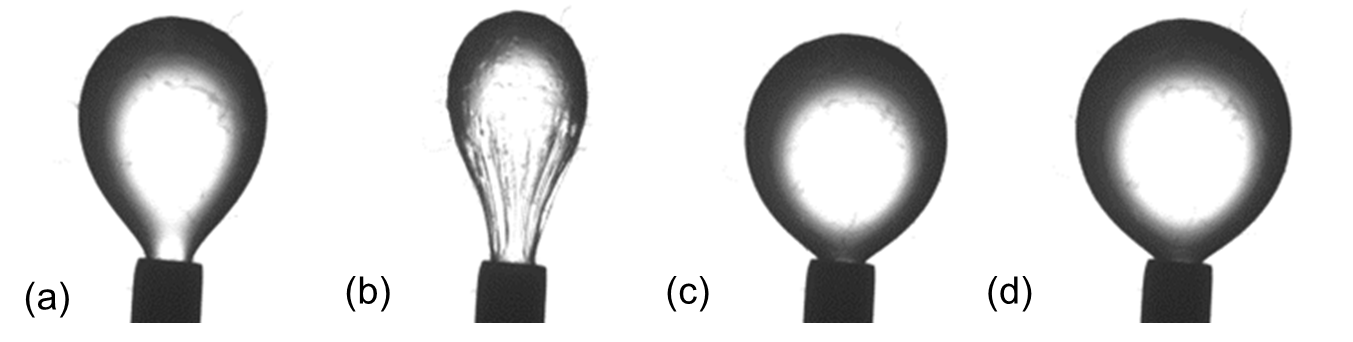}}\quad
\caption{\small{Compression and expansion of FBI on a drop of hexadecane, analysis performed with isotropic Young-Laplace equation; (a) initial drop, $\gamma_0 = 12.7~{\raise0.5ex\hbox{$\scriptstyle {mN}$}
\kern-0.1em/\kern-0.15em
\lower0.25ex\hbox{$\scriptstyle m$}}$, $V = 9.3~\mu l$, (b) drop after full compression, (c) inflation to initial drop size, more spherical, which is indicative of higher tension, $\gamma_0 = 21.5~{\raise0.5ex\hbox{$\scriptstyle {mN}$}
\kern-0.1em/\kern-0.15em
\lower0.25ex\hbox{$\scriptstyle m$}}$, $V = 9.3~\mu l$, (d) increase in drop size, increased tension, $\gamma_0 = 29~{\raise0.5ex\hbox{$\scriptstyle {mN}$}
\kern-0.1em/\kern-0.15em
\lower0.25ex\hbox{$\scriptstyle m$}}$, $V = 12.8~\mu l$. The needle is 0.9081 mm wide and serves as a scale bar.}}
 \label{SD11}
\end{figure*}
%%your .bib file
%\bibliography{ref-2} %your .bib file
%\bibliographystyle{rsc} %the RSC's .bst file
\end{document}